# Coexistence of near-$E_F$ van Hove singularity and in-gap topological Dirac surface states in superconducting electrides

Yin Yang [1,*], Peihan Sun [2,*], Ye Shen [1,*], Zhijun Tu [3,4,*], Pengcheng Ma [1], Hongrun Zhen [1], Tianqi Wang [1], Longli Tian [1], Tian Cui [1], Hechang Lei [3,4,†], Kai Liu [3,4,‡] and Zhonghao Liu [1,§]

[1] Institute of High-Pressure Physics and School of Physical Science and Technology, Ningbo University, Ningbo 315211, China
[2] Department of Physics, School of Science, Hebei University of Science and Technology, Shijiazhuang, Hebei 050018, China
[3] School of Physics and Beijing Key Laboratory of Optoelectronic Functional Materials & MicroNano Devices,
Renmin University of China, Beijing 100872, China
[4] Key Laboratory of Quantum State Construction and Manipulation (Ministry of Education), Renmin University of China, Beijing 100872, China

Superconducting electrides have attracted growing attention for their potential to achieve high superconducting transition temperatures ($T_C$) under pressure. However, many known electrides are chemically reactive and unstable, making high-quality single-crystal growth, characterization, and measurements difficult, and most do not exhibit superconductivity at ambient pressure. In contrast, La$_3$In stands out for its ambient-pressure superconductivity ($T_C \sim$ 9.4 K) and the availability of high-quality single crystals. Here, we investigate its low-energy electronic structure using angle-resolved photoemission spectroscopy and first-principles calculations. The bands near the Fermi energy ($E_F$) are mainly derived from La 5$d$ and In 5$p$ orbitals. A saddle point is directly observed at the Brillouin zone (BZ) boundary, while a three-dimensional van Hove singularity crosses $E_F$ at the BZ corner. First-principles calculations further reveal topological Dirac surface states within the bulk energy gap above $E_F$. The coexistence of a high density of states and in-gap topological surface states near $E_F$ suggests that La$_3$In offers a promising platform for tuning superconductivity and exploring possible topological superconducting phases through doping or external pressure.

Electrides are a unique class of materials in which electrons occupy spatially confined interstitial sites and act as anions. Unlike conventional anionic species, these interstitial electrons are not externally introduced but instead originate from the intrinsic electronic redistribution of the constituent atoms. Depending on the crystal structure and bonding environment, these electrons can form either localized or itinerant interstitial states [1–3]. Owing to their loose binding and spatial confinement, electrides exhibit a wide range of exotic physical and chemical properties, including low work function, high electron mobility, catalytic activity, and nonlinear optical responses [4–6]. More recently, electrides have attracted renewed attention for their potential to host superconductivity, especially under high pressure, where several compounds exhibit remarkably high superconducting transition temperatures ($T_C$) [7–9].

While many electrides are chemically reactive and unstable—making high-quality single-crystal growth challenging—the La$_3$X (X = Al, Ga, In, Tl, Sn, etc.) family crystallizing in the cubic Cu$_3$Au type, is notable for producing high-quality single crystals and exhibits superconductivity at ambient pressure [10, 11]. Within this series, La$_3$In shows the highest $T_C \sim$ 9.4 K—significantly exceeding that of elemental lanthanum [12–15]. Previous studies have suggested that strong electron–phonon coupling may play a dominant role in this enhancement [16]. The availability of well-ordered single crystals facilitates angle-resolved photoemission spectroscopy (ARPES) measurements of the detailed electronic structure, including high density of states (DOS) features such as van Hove singularities (VHSs) and saddle points (SPs)—both known to enhance many-body interactions and often linked to unconventional superconductivity and correlated phenomena [17, 18]. In La$_3$In, the coexistence of a VHS and the states from interstitial anionic electrons (IAEs) at the Fermi energy ($E_F$) may act synergistically to elevate $T_C$ and potentially induce strange-metal behavior [16]. Furthermore, the possible coexistence of topological surface states with superconductivity offers a promising platform for realizing topological superconductivity and exotic quasiparticles such as Majorana modes [19–24]. These unique characteristics underscore the importance of a comprehensive investigation of the electronic structure and surface states in this system.

In this work, we explore the low-energy electronic structure of single-crystalline La$_3$In using ARPES in combination with first-principles calculations. We identify a three-dimensional (3D) VHS at the Brillouin zone (BZ) corner that crosses $E_F$, as well as a SP located at the BZ boundary. Moreover, our calculations reveal the presence of topological Dirac surface states residing above $E_F$ within a bulk energy gap. These findings establish La$_3$In as a compelling platform for investigating the interplay among electron correlations, topological surface states, and superconductivity.

High-quality single crystals of La$_3$In were synthesized by the self-flux method as described elsewhere[16]. ARPES measurements were conducted at the 03U and Dreamline beamlines of the Shanghai Synchrotron Radiation Facility (SSRF). The energy and angular resolutions were better than 15 meV and 0.2°, respectively. Samples smaller than 1×1 mm² were cleaved *in situ*, yielding flat mirror-like (001) surfaces. During the measurements, the temperature was maintained at approximately 10 K, and the pressure was kept below $6.5\times10^{-11}$ Torr. The electronic structure calculations for La$_3$In were performed within the framework of density functional theory (DFT) using the projector augmented-wave (PAW) method [25, 26], as implemented in the Vienna *ab initio* Simulation Package (VASP) [27–29]. The exchange-correlation energy was treated using the generalized gradient approximation (GGA) in the Perdew-Burke-Ernzerhof (PBE) formulation [30]. A plane-wave energy cutoff of 520 eV was used throughout all calculations. The BZ was sampled using a $16 \times 16 \times 16$ Monkhorst-Pack $k$-point mesh. Both the lattice parameters and internal atomic coordinates were fully relaxed until the forces on each atom were less than 0.01 eV/Å and the total energy change was below $10^{-5}$ eV. After obtaining the optimized crystal structure, spin-orbit coupling (SOC) was included in the self-consistent electronic structure calculations. The surface states projected onto the two-dimensional (2D) BZ were computed using the WannierTools package [31], based on maximally localized Wannier functions constructed from the VASP outputs.

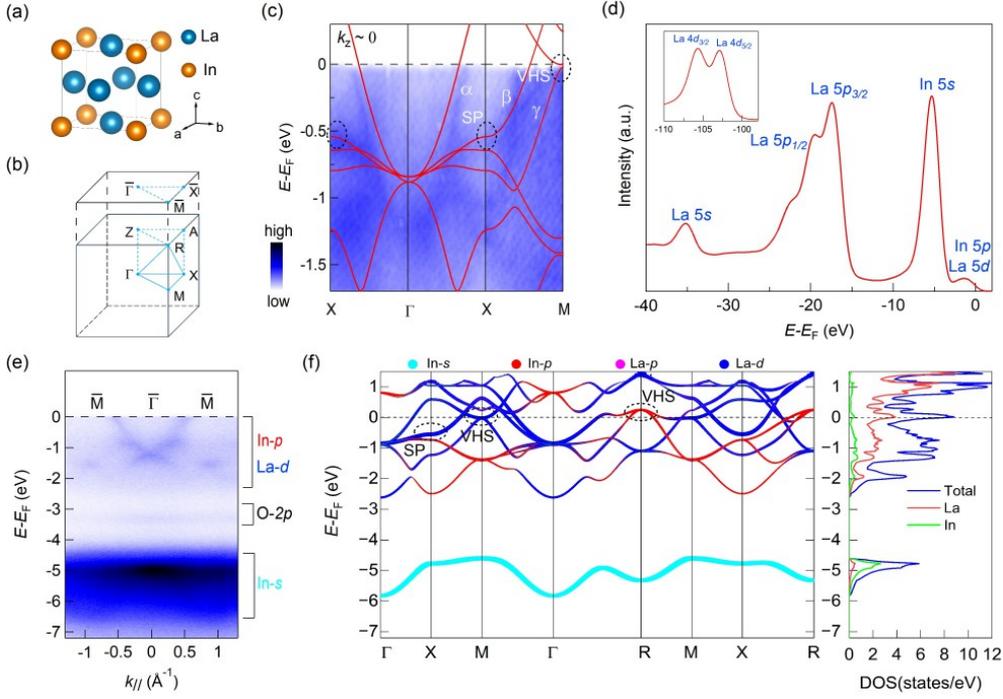

FIG. 1. (a) Crystal structure of La3In. (b) Bulk BZ and projected 2D BZ of the (001) surface. The indigo lines indicate the high-symmetry paths in the BZ. (c) Intensity plot along the Γ–X–M direction, overlaid with calculated bands (without renormalization or energy shift), showing the VHS at M and the SP at X. The black dashed ellipses indicate the SP and VHS. (d) Core-level photoemission spectrum displaying pronounced La and In peaks. Inset: Two La 4d peaks located about -105 eV below $E_F$. (e) ARPES intensity map along the $\overline{\Gamma}$–$\overline{M}$ direction, revealing the low-energy electronic structure; the dominant orbital contributions are indicated on the right. (f) Calculated orbital-projected band structure and DOS without SOC, where different colors denote distinct orbital characters.

The crystal structure of La$_3$In with space group $Pm-3m$ (No. 221) is depicted in Fig. 1(a). It can be described as a simple cubic lattice in which In atoms occupy the cube corners, while La atoms reside at the face centers, yielding a 3:1 stoichiometry (Cu$_3$Au type). Remarkably, replacing all In atoms with La yields elemental face-centered cubic Lanthanum, which is superconducting with $T_C \sim 5.5$ K [15]. The cubic BZ and its high-symmetry paths, including Γ–X–M–Γ–R–X and M–R, are illustrated in Fig. 1(b) with indigo solid lines. To facilitate comparison with photon-energy-dependent ARPES measurements, high-symmetry points Z and A on the $k_z \sim \pi$ plane are also marked, contrasting with Γ and X on the $k_z \sim 0$ plane. In Fig. 1(c), ARPES intensity plots along X–Γ–X–M are compared with calculated bands, revealing a SP at X located $\sim -0.5$ eV below $E_F$ and a VHS at M located exactly at $E_F$. Due to matrix-element effects, the hole-like band forming the SP at X shows asymmetric intensity along the two X–Γ branches; the left branch more clearly displays the hole-like dispersion, so both X points are shown here for clarity. Figure 1(d) presents the core-level photoemission spectrum, revealing sharp La 4d, 5s, 5p, 5d and In 5s, 5p peaks, indicating the excellent sample quality.

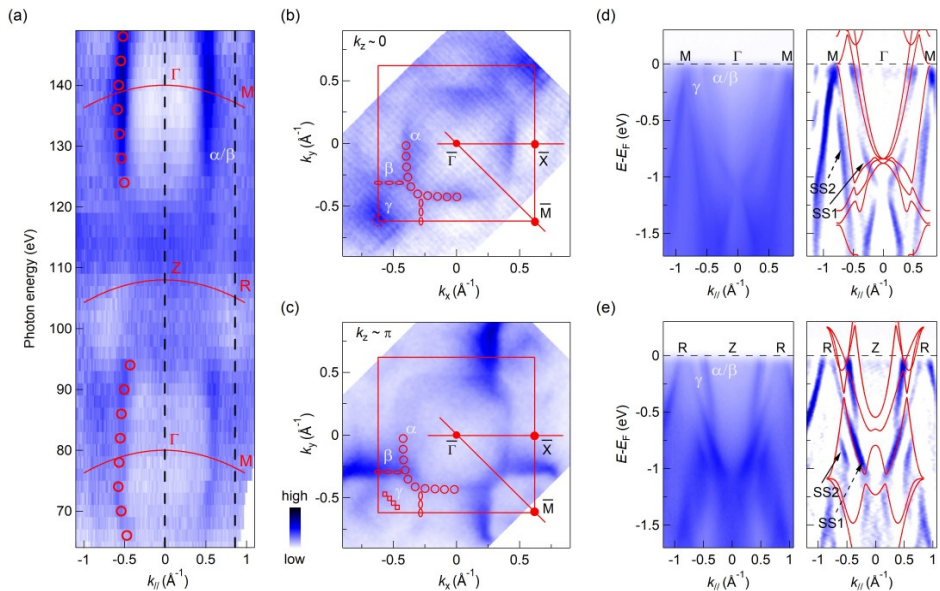

FIG. 2. (a) Integrated intensity map (±10 meV) on the $k_{//}$-$k_z$ plane at $E_F$. High-symmetry points and the $k_z$-dispersive bands are indicated. (b) Integrated intensity map at $E_F \pm 10$ meV taken on $k_z \sim 0$ plane. The red frames denote the 2D BZ projected onto the (001) surface, with high-symmetry points labeled. For guidance, parts of three Fermi surfaces (FSs) are outlined with red open circles. The $\gamma$ FS forms a hot spot at the BZ corner on the $k_z \sim 0$ plane. (c) Same as (d), but taken on the $k_z \sim \pi$ plane. The $\gamma$ FS is larger than that in panel (b), exhibiting a pronounced three-dimensional character. (d) Intensity plot and corresponding second-derivative plot along $\Gamma$–$M$ on the $k_z \sim 0$ plane, overlaid with the calculated bands. The Surface states (SS), SS1 and SS2, are indicated in the second-derivative plot. The solid arrows indicate the well-resolved surface states, while the dashed arrows mark where the corresponding surface states become mixed with the $k_z$-dispersive bulk states. (e) Same as (d), but along $Z$–$R$ on the $k_z \sim \pi$ plane.

The measured and calculated electronic structures over a wide binding-energy range are shown in Figs. 1(e) and 1(f), respectively. The orbital-projected band structure and DOS from first-principles calculations [Fig. 1(f)] are in good agreement with the experimental results [Fig. 1(e)] for the intrinsic La–In states, accurately reproducing the dispersions near $E_F$ derived from La 5$d$ and In 5$p$ orbitals. As indicated by the black dashed ellipses in Fig. 1(f), VHS and SP are predicted at the $M$ and $X$ points, respectively, contributing substantially to the DOS near $E_F$. Additionally, a flat band with weak intensity observed around $-3.3$ eV below $E_F$ in Fig. 1(e) is attributed to oxygen-derived states, consistent with oxygen atoms preferentially occupying IAEs sites in electrides [32]. Since oxygen is not included in the calculations, this feature is absent in Fig. 1(f). The localized interstitial states are likely formed through hybridization between nearby-energy orbitals of the intrinsic lattice atoms, leading to their distinct separation from the intrinsic La–In bands.

We performed photon-energy-dependent ARPES measurements to map the 3D low-energy band structure, as summarized in Fig. 2. Figure 2(a) displays the ARPES intensity at $E_F$ as a function of photon energy and inplane momentum $k_{//}$ along the $\Gamma$–$M$ ($Z$–$R$) direction. The $\alpha/\beta$ band, as well as the dispersion along $M$–$R$, exhibits a pronounced periodic modulation with photon energy in the range of 65–148 eV. Using an empirical inner potential of 13 eV and $c = 5.078$ Å, and applying the free-electron final-state model [33, 34], we determine that $h\nu = 80$ and 140 eV correspond approximately to the $\Gamma$ plane, whereas $h\nu = 108$ eV is close to the $Z$ plane.

Figures 2(b) and 2(c) show the Fermi surfaces (FSs) measured near the $k_z \sim 0$ and $k_z \sim \pi$ planes, respectively, revealing three sheets labeled $\alpha$, $\beta$, and $\gamma$. For clarity, the FS contours are marked by red open circles. Along the $\Gamma$–$M$ direction, the $\alpha$ and $\beta$ FSs remain difficult to distinguish across a wide photon-energy range due to the combined effects of finite resolution and $k_z$ broadening. In contrast, the $\gamma$ FS forms an intense hot-spot-like feature on the $k_z \sim 0$ plane and becomes significantly larger on the $k_z \sim \pi$ plane, demonstrating its pronounced 3D character. This behavior is further reflected in the energy–momentum dispersions along high-symmetry directions shown in Figs. 2(d) and 2(e). These data indicate that the $\alpha/\beta$ FSs originate from electron-like bands near $\Gamma$ and $Z$, whereas the $\gamma$ FS is derived from hole-like bands near $M$ and $R$. The VHS at $E_F$ arises from the top of the $\gamma$ band at $M$, but shifts above $E_F$ at $R$, showing that this band crosses the Fermi level along the $M$–$R$ direction. Such evolution is consistent with the strong $k_z$ dispersion of the high-DOS states along $M$–$R$ in Fig. 2(a).

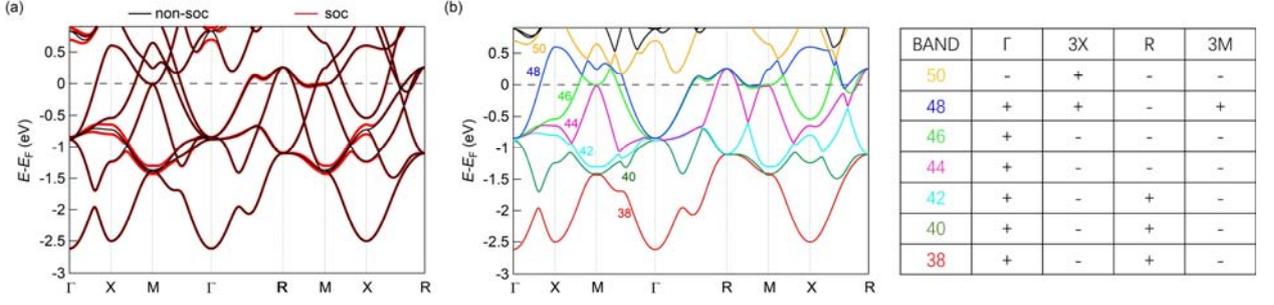

FIG. 3. (a) Calculated bands with SOC (red) overlaid with those without SOC (black). (b) The labeled calculated bands with SOC near $E_F$ distinguished by different colors. Parity products of the occupied states at the TRIM points are shown in the right table.

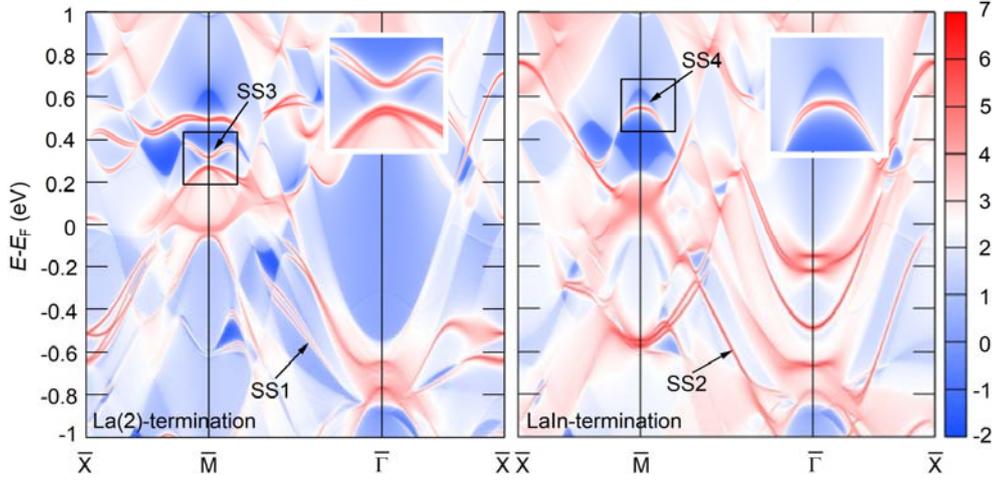

FIG. 4. Projected band structures for the La(2)- and LaIn-terminated (001) surfaces along the high-symmetry directions. The SS1 and SS2 correspond to observed surface states in Fig. 2. The Dirac SS3 and SS4 are highlighted and zoomed in.

When comparing the experimental dispersions with the calculated bands overlaid on the second-derivative spectra, we find excellent agreement on the $k_z \sim 0$ plane. However, the calculated bands appear more intricate near $E_F$ on the $k_z \sim \pi$ plane, likely due to band entanglement originating from correlation effects and interband interactions, leading to certain discrepancies from the experimental data. Additionally, two extra features labeled SS1 and SS2 in Figs. 2(d) and 2(e) correspond to surface states, as confirmed by comparison with the calculated results in Fig. 4. The observed surface states contain contributions from both calculated modes: SS1 (associated with the La(2) termination) is clearly visible near the $k_z \sim 0$ plane but merges with bulk states near $k_z \sim \pi$, whereas SS2 shows the opposite trend. These behaviors indicate that the measurements involve a mixture of the two terminations and that the bulk states are interwoven with these surface states.

Figure 3(a) compares the electronic structures calculated with and without SOC. Since La$_3$In preserves both inversion and time-reversal symmetries, Kramers degeneracy ensures that all bands remain doubly degenerate. As a result, SOC does not induce significant band splittings near the $E_F$, apart from small symmetry-allowed gaps at band crossings. Moreover, according to the Bohr model, the strength of SOC is proportional to the fourth power of the atomic number and inversely proportional to the azimuthal quantum number $l$ (except for $l = 0$) [35]. In the present case, the low-energy states are mainly derived from In-5$p$ and La-5$d$ orbitals (Fig. 1(f)). The 5$p$ orbitals of In primarily contribute regions exhibiting larger SOC-induced band splittings, while those with smaller splittings originate mainly from the 5$d$ orbitals of La. Although La has a larger atomic number than In, the $d$ orbitals ($l = 2$) possess a higher azimuthal quantum number than the $p$ orbitals ($l = 1$). Consequently, under the combined effect of these factors, the SOC-induced band splitting is more pronounced in the In-dominated bands than in the La-dominated bands.

As shown in Fig. 3(b), once the SOC is taken into account, the band crossings between the 48th and 50th bands around the $M$ point are fully gapped, yielding a continuous gap throughout the entire BZ. This is because the bands of La$_3$In along the $X-M$, $M-\Gamma$, and $X-R$ directions possess $C_{2v}$ double point group symmetry, which cannot protect the band crossings along those high-symmetry paths [36]. In this sense, superconducting La$_3$In can be regarded as an insulator defined at a curved Fermi level between the 48th band (blue) and the 50th band (orange). Owing to the coexistence of time-reversal and inversion symmetries, its topological invariant $\mathbb{Z}_2$ can be determined from the product of the parity eigenvalues of all occupied bands at the eight TRIM points [37], namely, $\Gamma$, three $X$, $R$, and three $M$ points [Fig. 1(b)]. As listed in the table, the resulting $\mathbb{Z}_2 = 1$, confirming the nontrivial topological character of La$_3$In.

Besides the $\mathbb{Z}_2$ topological invariant, the nontrivial topology of La$_3$In is directly manifested in its surface states. We calculated the (001) surface spectrum, considering both La(2) and LaIn terminations, as shown in Fig. 4. The SS1 and SS2 are experimentally observed as mentioned above. The robust Dirac surface states SS3 and SS4, protected by time-reversal symmetry, appear clearly within the bulk gap, confirming their topological origin. Specifically, SS3 corresponds to the La(2) termination, where an in-gap Dirac cone emerges at the $M$ point about 0.30 eV above $E_F$, while SS4 corresponds to the LaIn termination, with the Dirac cone located near 0.55 eV above $E_F$. These states are well disentangled from bulk bands, reflecting their symmetry-protected and termination-resilient nature, and the different energy positions highlight surface-dependent tunability. Notably, the Dirac states at $M$ lie close to a 3D VHS with a high DOS. Given that La$_3$In is a conventional BCS superconductor, shifting the chemical potential via doping, gating, or strain could change electronic correlations and drive interesting quantum phase transitions. In concert with the nontrivial surface states, such tuning may also provide a pathway to topological superconductivity.

In summary, we have investigated the low-energy electronic structure of the superconducting electride La$_3$In by high-resolution ARPES and first-principles calculations. We identified La-5$d$ and In-5$p$ dominated high-DOS states near $E_F$, including a saddle point at $-0.5$ eV and a 3D VHS crossing $E_F$, which likely enhances correlations and the superconducting transition temperature. Parity analysis at the TRIM points yields a nonzero $\mathbb{Z}_2$ invariant, confirming nontrivial topology. The resulting Dirac surface states are symmetry-protected, robust, and well disentangled from bulk bands, with slightly different energies for La(2) and LaIn terminations, enabling surface-dependent tunability. These clean in-gap modes provide a promising platform for exotic transport and device applications. Our findings establish La$_3$In as a rare system in which superconductivity, a 3D VHS, and topological surface states coexist, offering unique opportunities to explore the interplay among topology, strong correlations, and superconductivity in electride materials.

This work was supported by the National Natural Science Foundation of China (NSFC, Grants No. 12222413, 12174443, 12274459, and 12404266), the National Key R&D Program of China (Grants No. 2023YFA1406500 and 2022YFA1403800), the Natural Science Foundation of Shanghai (Grants No. 23ZR1482200), the Natural Science Foundation of Ningbo (Grants No. 2024J019), the Science Research Project of Hebei Education Department (Grant No. BJ2025060), the funding of Ningbo Yongjiang Talent Program and the Mechanics Interdisciplinary Fund for Outstanding Young Scholars of Ningbo University (Grants No. LJ2024003). We thank the SSRF of BL03U (31124.02.SSRF.BL03U) for the assistance with ARPES measurements. Computational resources were provided by the Physical Laboratory of High-Performance Computing at Renmin University of China and the Beijing Super Cloud Computing Center.

* Equal contributions
† hlei@ruc.edu.cn
‡ kliu@ruc.edu.cn
§ liuzhonghao@nbu.edu.cn